# What are the Sources of Solar Energetic Particles?
## Element Abundances and Source Plasma Temperatures

Donald V. Reames

*IPST, Univ. of Maryland, College Park, MD*

**Abstract** We have spent 50 years in heated discussion over which populations of solar energetic particles (SEPs) are accelerated at flares and which by shock waves driven out from the Sun by coronal mass ejections (CMEs). The association of the large "gradual" SEP events with shock acceleration is supported by the extensive spatial distribution of SEPs and by the delayed acceleration of the particles. Recent STEREO observations have begun to show that the particle onset times correspond to the observed time of arrival of the shock on the observer's magnetic flux tube and that the SEP intensities are related to the local shock speed. The relative abundances of the elements in these gradual events are a measure of those in the ambient solar corona, differing from those in the photosphere by a widely-observed function of the first ionization potential (FIP) of the elements. SEP events we call "impulsive", the traditional "$^3$He-rich" events with enhanced heavy-element abundances, are associated with type III radio bursts, flares, and narrow CMEs; they selectively populate flux tubes that thread a localized source, and they are fit to new particle-in-cell models of magnetic reconnection on open field lines as found in solar jets. These models help explain the strong enhancements seen in heavy elements as a power (of $2-8$) in the mass-to-charge ratio $A/Q$ throughout the periodic table from He to Pb. A study of the temperature dependence of $A/Q$ shows that the source plasma in impulsive SEP events must lie in the range of 2–4 MK to explain the pattern of abundances. This is much lower than the temperatures of >10 MK seen on closed loops in solar flares. Recent studies of $A/Q$-dependent enhancements or suppressions from scattering during transport show source plasma temperatures in gradual SEP events to be 0.8–1.6 MK in 69% of the events, *i.e.* coronal plasma; 24% of the events show reaccelerated impulsive-event material.

## 1 Introduction

It seems like an easy question.  Do solar energetic particles (SEPs) originate suddenly at the relative point source of a solar flare or at the extended, continuing source of a CME-driven shock wave?  The answer: "both", only increases the difficulty.

There is clearly shock acceleration.  Shock waves passing near Earth show peak intensities at the shock extending to protons at >700 MeV (see Figure 13 in Reames 2013).  These shocks are much stronger nearer the Sun, at 2–3 solar radii (Reames 2009b), than near Earth and must contribute more in transit from the Sun to the particles seen earlier and at even higher energies.

There is also evidence of an acceleration mechanism that enhances $^3$He/$^4$He in small SEPs by several orders of magnitude and enhances heavy elements by similar amounts (see reviews Reames 1999, 2013; Mason 2007).   This mechanism cannot be a shock, although subsequent shock acceleration is not excluded.

There must be at least two mechanisms, but are there more?  This has certainly complicated the history of the subject, which at times is even emotional.  The reaction was hostile to the paper "The Solar Flare Myth" by Gosling (1993) which simply reviewed the production of big SEP events by CME-driven shocks rather than by flares.  The controversy led to an invited discussion from three alternative viewpoints in Eos where Hudson (1995) argued that "flare" should include the CME, shock, and any related physics, Miller (1995) argued that flares, being more numerous, were a better subject for acceleration studies, and Reames (1995b) argued for the separate study of both flare and shock acceleration of SEPs (see also Hudson *et al.* 1995; Svestka 1995; Reames 1995c).

We begin with a brief history of early ideas about SEP sources and considerations affecting SEP origin, in section 2 we discuss more-recent evidence, and in section 3 we discuss broader evidence from element abundances.

### 1.1 Radio Emission

The frequency of radio emission depends upon the electron density of the excited plasma which decreases with distance from the Sun.  Fast-drift type III bursts are produced when 10–100 keV electrons excite the plasma as they stream out from the source near the Sun and slow-drift type II emission is produced at a shock wave moving outward more slowly at a speed of the order of 1000 km sec$^{-1}$.  Fifty years ago, Wild Smerd, and Weiss (1963) suggested that electrons were accelerated to produce the type III emission while protons were accelerated at shock waves associated with type II bursts.  This already implied *two* distinct sources of the solar energetic particles we observe.

### 1.2 Coronal Diffusion and the Birdcage Model.

However, the ideas from the radio observations were ignored by a large segment of the SEP community.   They assumed that the particles must be accelerated by *some* unspecified mechanism at a flare.  The wide latitude spread observed in large SEP events was thus treated as diffusion of particles across the corona from the flare followed by diffusion out along the magnetic field (Reid 1964).  Coronal loops, forming structures like a birdcage, spread the particles which passed across the corona in a diffusion-like process from loop to loop by magnetic reconnection at the loop footpoints (Newkirk and Wenzel 1978). Diffusion models have been helpful in understanding particle acceleration and transport but the actual physics of the underlying scattering process is contained within the diffusion coefficient. Unfortunately the diffusion coefficients are sometimes treated as adjustable parameters and their relation to physical processes becomes obscured.

These models held sway for a decade. The International Cosmic Ray Conferences held entire sessions on Coronal Transport up to 1990. Most of the papers in these sessions involved diffusion calculations.

The birdcage model would surely cause strong longitude variations in the element abundances and rigidity spectra, especially below 1 MeV amu$^{-1}$. Mason, Gloeckler and Hovestadt (1984) pointed out that such variations with solar longitude were simply not observed and they suggested large-scale shock acceleration to account for the observed particles. Subsequently, Cane, Reames, and von Rosenvinge (1988) showed that the time evolution of energetic protons as a function of solar longitude fit the expected pattern for shock acceleration in a magnetic field swept eastward into the Parker spiral.

### 1.3 Big Flare Syndrome

In 1982, Kahler (1982) pointed out the dangers of correlations produced by "big flare syndrome." Big flares are often accompanied by big CMEs and big SEP events, producing correlations that do not necessarily show a causal relationship. A major reconnection in the solar atmosphere may cause larger flares and CMEs, producing a correlation between them. That does not mean that flares *cause* CMEs or the converse. It is easy to be misled by big flare syndrome, and many have been.

Outside of these correlations there are also examples of poor correlations. Disappearing filament events produce CMEs and SEP events with no flares (Kahler et al 1986; Cane et al 1986) or with weak flares (Gopalswamy *et al.* 2015) and in October 2014 a series of X-class flares from E64 to W57 produced weak or absent CMEs and no energetic particles at all (Tylka, private communication, but see Thalmann *et al.* 2014).

### 1.4 SEP Association with CME-driven Shocks

A relationship between SEP events and CMEs began in earnest with the evidence that 96% of large SEP events were associated with fast, wide CMEs (Kahler *et al.* 1984). These long duration or "gradual" SEP events were to become associated with the large-scale shock acceleration of Mason, Gloeckler, and Hovestadt (1984) and the type II bursts of Wild, Smerd and Weiss (1963). High CME speeds could be measured even when type II bursts were not generated. Furthermore, there was evidence at this time that the SEPs formed a continuous spectral distribution extending from energies just above the solar wind (Gosling *et al.* 1981); the SEPs in big events were just accelerated solar wind.

At about the same time, Lee (1983) adapted shock theory to an interplanetary setting. In this theory, particles were accelerated as they scattered back and forth across the shock while protons streaming away from the shock amplified the waves that produced the scattering which then limited and balanced the streaming. Subsequently, this work has inspired modeling of fully time-dependent shock acceleration of protons to hundreds of MeV with self-generated turbulence (*e.g.* Ng and Reames 2008).

### 1.5 He-rich SEP Events

The ratio of $^3$He/$^4$He in the solar wind is $5 \times 10^{-4}$ (*e.g.* Coplan et al. 1985). Hsieh and Simpson (1970) first noticed order of magnitude enhancements of $^3$He/$^4$He in SEP events. Subsequent measurements even found $^3$He/$^4$He >10 so that the intensity of $^3$He exceeded that of H in a few events (Reames, von Rosenvinge, and Lin 1985). These $^3$He-rich events were associated with electrons and with type III radio bursts (Reames and Stone 1986). $^3$He-rich events were accompanied by enhancements in heavy elements up to a factor of ~10 in Fe/O (see reviews by Mason 2007; Reames 1999, 2013).

Clearly, the $^3$He enhancements could not be produced by shock acceleration like the gradual events. These "impulsive" events, named for their typically shorter durations, must involve a second acceleration mechanism. A highly resonant wave particle interaction seemed likely; eventually one with waves produced by streaming type III electrons and preferentially absorbed by $^3$He (Temerin and Roth 1992; Roth and Temerin 1997).

### 1.6 Reacceleration

For a time, there seemed to be a clear separation between gradual and impulsive events and their distinct acceleration mechanisms based partly upon element and isotopic abundances. However, Mason, Mazur and Dwyer (1999) found small enhancements of $^3$He, by a factor of only 5, in large "gradual" SEP events where they were not expected. The authors understood the $^3$He to result from shock acceleration of suprathermal ions left over from previous impulsive SEP events.

Shock re-acceleration of superthermal ions with differing source abundances and ionization states has become an important factor in understanding the energy spectra and abundance variations that are observed in large SEP events (Desai *et al.* 2004; Tylka *et al.* 2005; Desai *et al.* 2006). High-energy breaks in the energy spectra that depend on the mass-to-charge ratio *A/Q* of the ions as well as the angle between the magnetic field and the shock normal $\theta_{Bn}$, can produce complex abundance patterns, usually above about 10 MeV amu$^{-1}$ (Tylka and Lee 2006).

As shock waves move outward from the Sun they continue to accelerate particles to MeV energies, sometimes hundreds of MeV, from a varied background of particles, especially those accelerated earlier by the same shock. Several authors have suggested that particle acceleration is enhanced by the presence of a previous CME or SEP event (Kahler 2001; Gopalswamy *et al.* 2002, 2004; Cliver 2006; Li *et al.* 2012; Reames 2013). Reacceleration by shock waves is common and it complicates the spectra and abundances in SEP events.

## 2 More-Recent Evidence of SEP Origin

New kinds of observational evidence for the origin of SEPs have continued to emerge since the early days. We present some of that evidence in this section.

### 2.1 Evidence of a Compact Source

Particles emitted from a highly localized source near the Sun will emerge on a limited number of magnetic flux tubes that thread the source. Random walk of magnetic field lines will have previously mixed flux tubes that are to be suddenly filled if they penetrate the compact SEP source or else they remain empty when each flux tube is convected past an observer in space. This is the interpretation of the observations to the left in Figure 1 from Mazur *et al.* (2000). Particles of decreasing energy arrive at later times, but clear time gaps are seen for the impulsive events in the upper left panel corresponding to the arrival of magnetic flux tubes that do not thread the particle source at the Sun. All flux tubes are filled for the extended shock source of the gradual SEP event on the right.

More recent studies by Chollet and Giacalone (2011) show that the boundaries between the flux tubes with and without SEPs in the impulsive events are extremely sharp, indicating that there is no evidence of cross-field propagation during particle transit to 1 AU. These particles may eventually find paths across to other flux tubes; impulsive events are occasionally seen at distant longitudes from their source flares, but the time scales for this transit are long and the intensity attenuation is significant.

In any case, the intensity dropouts in Figure 1 are not only evidence of confinement of ions to magnetic flux tubes, but are evidence of compact sources for impulsive SEP events. In contrast, the lack of such dropouts in gradual events shows evidence of a spatially extensive source.

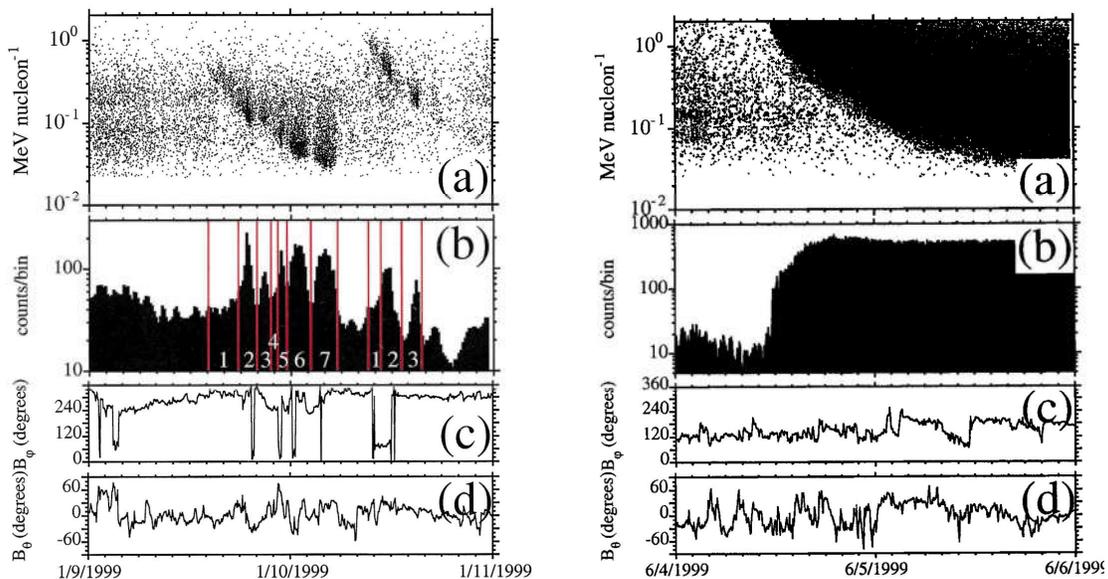

**Fig. 1** Panels (a) plot the energy *vs.* arrival time for individual ions from an impulsive (left) and gradual (right) SEP event. Panels (b) show the corresponding ion count rates while (c) and (d) show the magnetic field direction (Mazur *et al.* 2000).

## 2.2 Improving SEP-Shock Correlations

How can we find out if the correlations between SEP intensities and CME speeds discussed by Kahler (2001) are not merely "big flare syndrome" from Kahler (1982). First we must recognize that there is no single SEP intensity in an event since intensity varies with longitude (Reames, Kahler, and Ng 1997; Lario 2013). Also, there is no single speed of the CME-driven shock, which also varies with space and time across a broad front. If SEPs are truly accelerated at CME-driven shocks, we should be able to improve the correlations by being more careful about which region on the shock might accelerate the particles we measure, and when.

Image data from the STEREO spacecraft allows a mapping of both the CME and shock, and a reconstruction of their evolution in time as shown in Figure 2 (Rouillard *et al.* 2011, 2012). In some cases this allows a correlation of the SEP onset time with the time the shock strikes the observer's field line, later on the flanks of the CME than at its nose (see Reames 2009b; Gopalswamy *et al.* 2013). In other cases the SEP peak intensities seen by different observers can be correlated with the shock speeds at the base of their field lines. Such correlations are shown in red on a background of earlier event-averaged correlation plots from Kahler (2001) in the lower right panels of Figure 2.

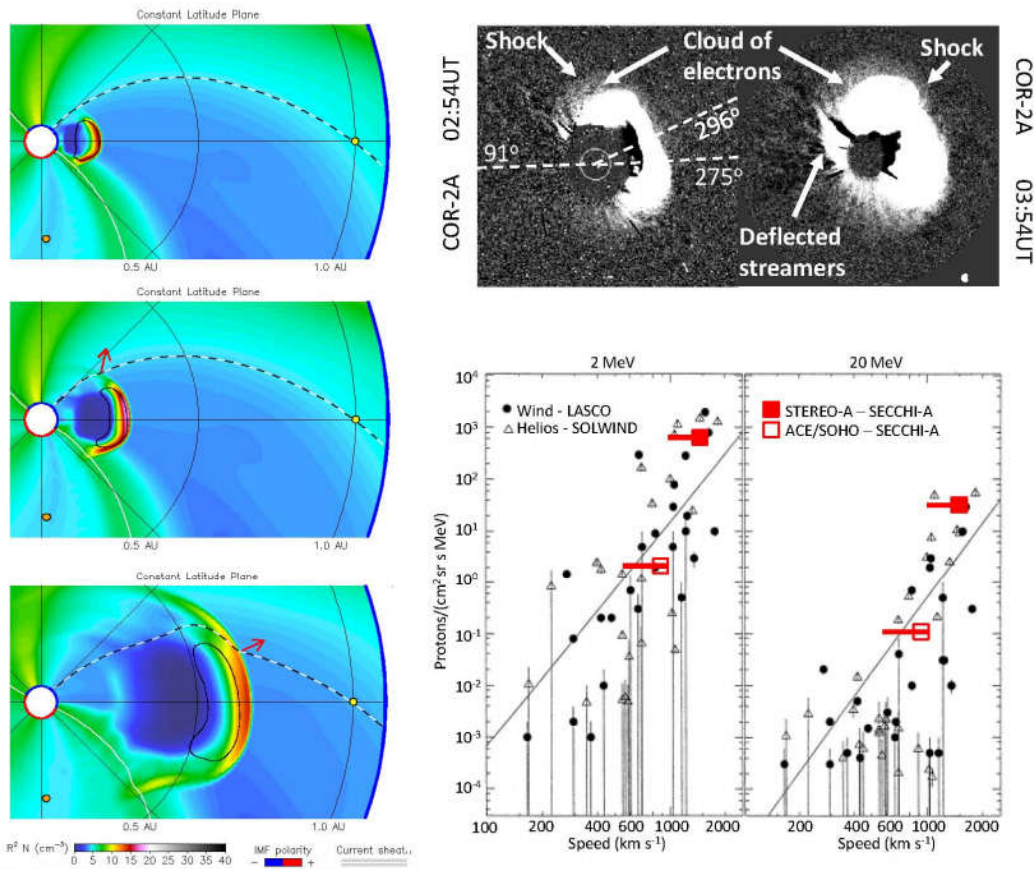

**Fig. 2**. Coronagraphs on the STEREO spacecraft can image CME-driven shock waves, as shown in the upper right panels, allowing reconstruction of CME and shock evolution as shown along the left-hand panels. This allows timing of the expected SEP acceleration at shock contact and correlations on SEP intensity *vs.* shock speed for different spacecraft shown on a background of previous correlations for 2 and 20 MeV protons in the lower right panels (Rouillard *et al.* 2011, 2012).

The correlations and onset timing seem to be improved in these studies, but more examples would be helpful. If the correlations improve, shock acceleration is supported, if not, we have only big flare syndrome. In principle, one could model the SEP acceleration and transport as a function of time produced by an observed and varying shock along any given flux tube.

## 2.3 Onset Timing

Comparison of the timing of the particle release at the Sun with X-ray or γ-ray production at the associated flare is an important test of their physical association. When relativistic protons can be measured in the ground-level events (GLEs) produced by secondary neutrons from nuclear reactions of the primary protons in the atmosphere, the proton transit times to Earth are only slightly greater than those of the photons. Unfortunately, most GLEs rise less than 10% above the background produced by galactic cosmic rays and clear onsets of GeV protons in neutron monitors are only available for a few large GLE events. However, we can use the velocity dispersion of the ions of lower energy to study a larger number of events.

As shown in the left panel of Figure 3, higher velocity ions arrive earlier in a clear progression so that the transit time of the first arriving ions equals their path length from the Sun divided by the ion velocity *v*. A fit of the onset times *vs.* 1/*v* determines the solar-particle (SPR) release time and the path length along the magnetic field. Theoretical studies show that scattering only delays the earliest particles by 2-3 min (see appendix in Rouillard *et al.* 2012). The earliest arriving particles have been focused to travel along the diverging magnetic field.

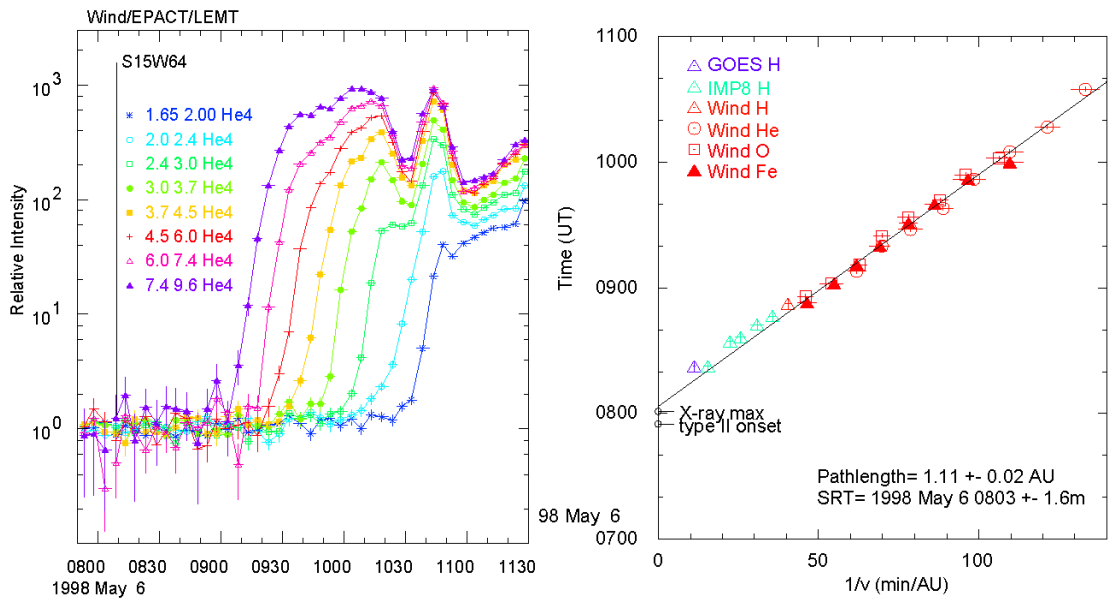

**Fig 3.** Ions of different energy arrive in inverse velocity order in the 1998 May 6 SEP event (left panel). A linear fit of the onset times *vs.* 1/*v* shown in the right panel determines the solar release time as the intercept and the path length along the field as the slope (Reames 2009a, b).

The SPR times from studies of this kind from impulsive ($^3$He-rich) SEP events and gradual events (GLEs) have been compared with photon data by Tylka *et al.* (2003) as shown in Figure 4. SPR times correspond well with hard X-ray peaks for the impulsive events but in GLEs they lag the $\gamma$-ray peaks by up to half an hour.

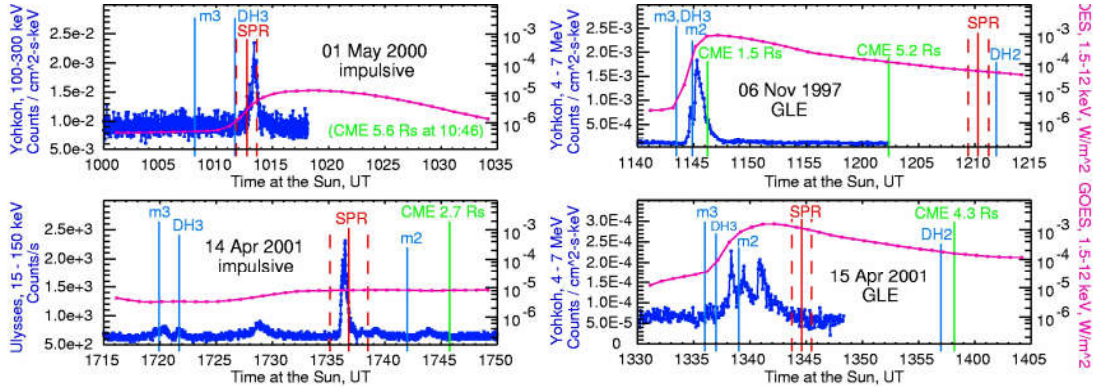

**Fig.4.** Solar particle release (SPR) times (red) are compared with hard X-ray profiles (blue) for impulsive SEP events in the two left panels and with $\gamma$- ray line intensities (blue) for GLE events in the right panels. Times of radio bursts (light blue) and CME observations (green) are shown as are soft X-ray profiles (magenta). From Tylka *et al.* (2003).

It is sometimes naively suggested that the delayed SPR times occur because SEPs are stored on magnetic loops near the Sun then later released. However, storage near the Sun is just what led to the downfall of the birdcage model. It would lead to substantial modification of the spectra and abundances of low-energy ions in GLEs, which are not observed. In fact, about half of the time periods examined by Mason, Gloeckler, and Hovestadt 1984) occurred during GLEs. Particles that are actually accelerated and stored on closed loops are soon scattered into the loss cone and plunge into the lower corona to generate $\gamma$-rays in the energy region of nuclear $\gamma$-ray lines (4–7 MeV) that are observed as the dark blue profiles in the right-hand panels in Figure 4. The particles seen in space begin their journey well after the nuclear $\gamma$-rays have decayed away in these GLEs.

In a shock model we would expect the shock to intercept field lines and accelerate particles later on the flanks of the CME than at its nose (see Reames 2009b; Gopalswamy *et al.* 2013). Such behavior was shown in the left panels of Figure 2. Perhaps we should divide the GLEs into two categories; "late" which surely must be shock accelerated and "early" which could come from either flare or shock based upon timing alone. However, the early GLEs also have fast, wide CMEs just like the late ones (Gopalswamy *et al.* 2012), so possibly the shock for these just forms earlier and lower in the corona. We need no flare acceleration at all to produce the particles seen in space. In fact, events with earlier SPR times may only indicate the presence of faster shock waves. While SPR times have been measured for many GLEs (*e.g.* Reames 2009a, b), we need $\gamma$-ray line intensities in more events for better comparisons.

The path lengths and timing of non-relativistic (*e.g.* 27 keV) electrons are in agreement with those found for ions in GLEs and they may come from the same shock source (Tan *et al.* 2013). Relativistic electrons are delayed by scattering while non-relativistic electrons travel scatter-free (Tan *et al.* 2011).

## 2.4 Injection Profiles

Given the time dependence of the particle intensities and the height-time plot for CMEs, Kahler (1994) constructed injection profiles of intensity *vs.* CME height as shown in Figure 5. These show significant acceleration of GeV protons only when the CMEs reach heights above about 5 solar radii.

**Fig. 5.** Injection profiles of energetic protons *vs.* CME height are shown for three GLEs: 1989 August 16, 1989 September 29, and 1989 October 24 (Kahler 1994).

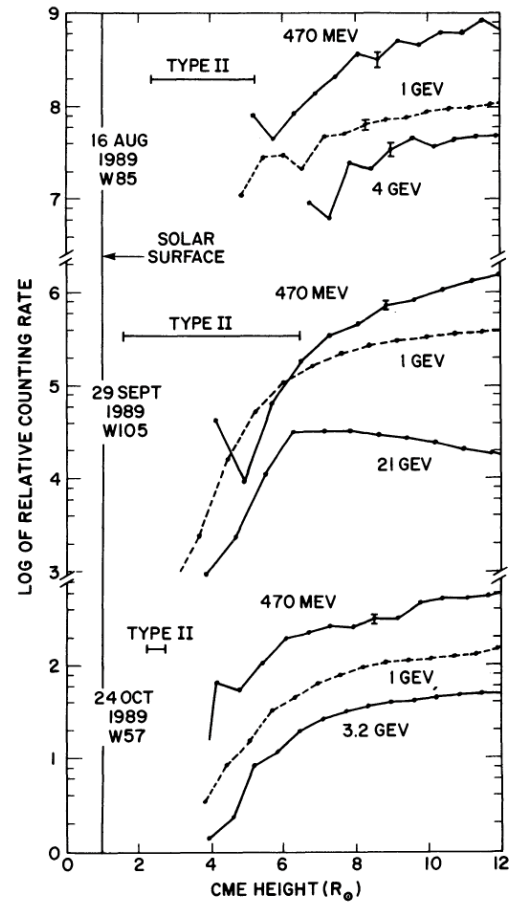

## 2.5. Energy Spectra

Much of the evidence used to associate large gradual SEP events with shock acceleration, especially abundances, comes from observations at intermediate energies, below 100 MeV amu$^{-1}$. Is it possible that particles at these low energies come from shocks, but the protons above 1 GeV come from flares?

The left-hand panel in Figure 6 shows a comparison of the proton and He spectra from 1 MeV amu$^{-1}$ to 10 GeV amu$^{-1}$ from Lovell, Duldig, and Humble (1998). Spectra at high energies (shown as shading) are deduced using the asymptotic look directions and cutoff rigidities of the neutron monitor network to determine proton spectra and angular distribution. The fit, using the shock model of Ellison and Ramaty (1985), shows that the proton spectrum rolls downward in this very large event.

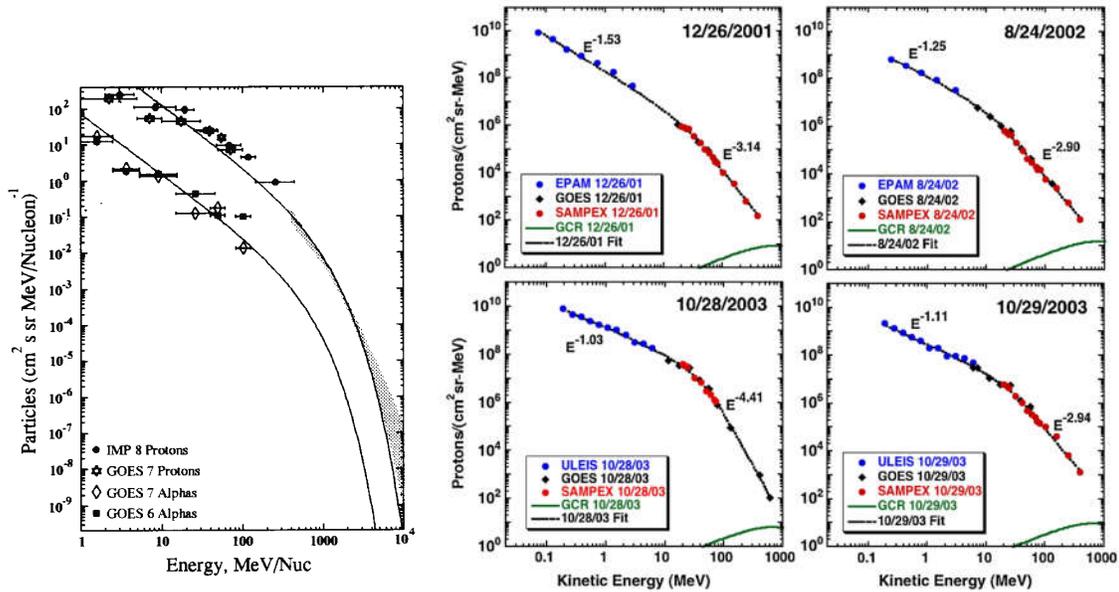

**Fig. 6.** The left-hand panel shows energy spectra in the large GLE event of 1989 September 29 (Lovell, Duldig, and Humble 1998). The shaded region at high energies is the spectrum deduced from neutron monitor measurements; the fitted curves are shock spectral forms from Ellison and Ramaty (1985). The four right-hand panels are GLE proton fluence spectra from Mewaldt *et al.* (2012).

The four right-hand panels in Figure 6 are a sample of the 16 GLE spectra from Mewaldt *et al.* (2012) who also summarized the properties of the energy spectra and element abundances in GLEs. None of the GLEs show evidence of spectral flattening at high energy that might signal a new (flare) component with a harder spectrum. However, the element abundances at high energies sometimes reflect the properties of impulsive suprathermal ions from the seed population swept up by the shock which are expected theoretically to appear at high energies (Tylka and Lee 2006). GLEs only differ from other gradual events in having slightly more intense GeV proton spectra – an extreme tip of a continuous distribution.

Following a suggestion that acceleration may occur at the magnetic reconnection of coronal arcades following CMEs, Kahler, McAllister, and Cane (2000) found 30 arcades with no detectible SEP increases. Five arcades did show SEP increases that were probably accelerated by shocks that were present.

## 3 Element Abundances

Element abundances are a wider topic that goes beyond simply distinguishing the particle source. The events we identify as gradual provide a measure of the element abundances in the corona itself. Those we classify as impulsive appear to give us a measure of the temperature of the source plasma in magnetic reconnection regions on open field lines near solar flares and physical parameters of that acceleration. While we generally refer to abundances of elements, electron abundances also serve to distinguish impulsive and gradual events (Cliver and Ling 2007, 2009).

Figure 7 shows a comparison of the element intensities (lower panel) and relative abundances (upper panel) of an impulsive SEP event followed by a gradual event in April of 2001. Intensities of the 20 MeV protons and the C barely exceed the background at the onset of the impulsive event, but Ne and Fe rise by orders of magnitude. The GLE is clearly noted by a rise of several orders of magnitude in intensities of all species.

**Fig. 7.** Intensities of 20 MeV protons and of C, Ne, and Fe at 3–5 MeV amu⁻¹ are shown in the lower panel. Ratios of Fe/C and Ne/O, normalized to coronal abundances, are shown in the upper panel. Onset times of an impulsive event and a gradual GLE event are flagged with the solar latitude and longitude of the source (Reames 2014).

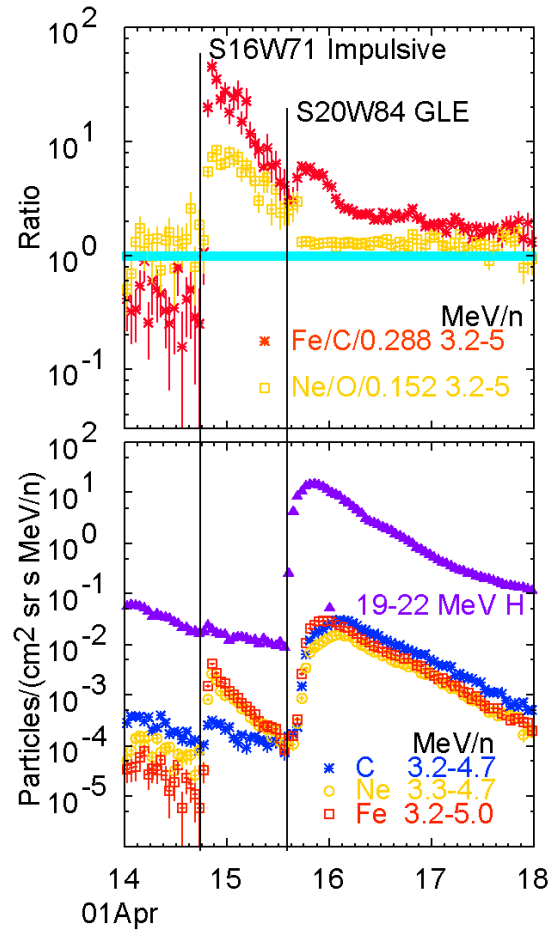

The abundance ratios Fe/C and Ne/O shown in Figure 7 are quite high in the impulsive event. They decrease with time partly because the intensities of C and O approach background and cannot follow the decrease in Fe and Ne intensities. Fe/C is elevated in the gradual event but decreases toward the coronal abundance with time. Ne/O is constant throughout the gradual event.

We can most efficiently find impulsive events by scanning for periods of enhanced Fe/O or Ne/O. Figure 8 shows a two-dimensional histogram of enhancements in Ne/O *vs.* Fe/O for all 8-hr periods observed on the Wind spacecraft with well-defined abundance measurements in the selected energy intervals (Reames, Cliver, and Kahler 2014a). We use Fe/O rather than ³He/⁴He to distinguish impulsive events since ³He/⁴He has large variations with energy (*e.g.* Liu, Petrosian, and Mason 2006) and is thus poorly defined. Furthermore, if we wish to measure element abundances we need periods with adequate statistics of Fe and O.

**Fig. 8.** A histogram of 8-hr periods with enhancements of Ne/O *vs.* Fe/O shows a peak for gradual events around (1,1) and for impulsive events beginning near (6, 3) (Reames, Cliver, and Kahler 2014a).

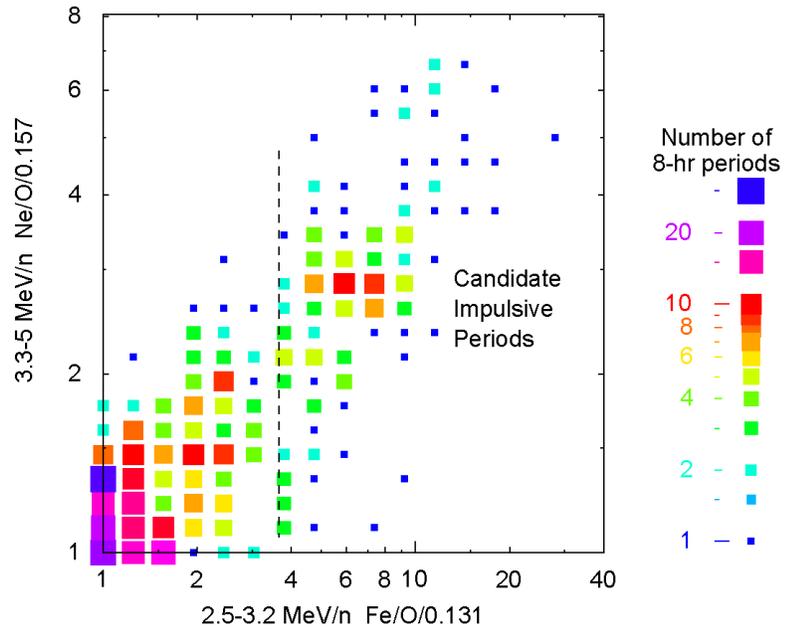

## 3.1 Gradual Events – Averaged Abundances

Particles streaming out along magnetic field lines generate Alfvén waves that scatter those coming behind. This is especially important for intense gradual events (Reames, Ng, and Tylka, 2000; Ng, Reames, and Tylka 2003; Reames and Ng 2010). Approximately, if the frequency spectrum of scattering waves is a power law, the scattering mean free path $\lambda$ is a power of particle rigidity, so, if we compare ions of the same velocity, $\lambda \sim (A/Q)^a$. Since $A/Q$ for Fe exceeds that of C, or O, Fe will usually scatter less and arrive early, so that Fe/C rises initially and decreases with time, while Ne/O, involving species with similar values of $A/Q$, will vary little with time. This behavior is seen for the gradual event in Figure 7. If we wish to recover a "coronal" abundance sample accelerated at the shock, we must average over the spatial fractionation produced by differential scattering during transport.

Figure 9 shows this behavior for three large SEP events where the scattering is sufficient to produce Fe-enhanced regions early in the events followed by Fe-depleted regions later. For less intense events, the scattering is reduced and the Parker spiral field tends to mold this into Fe enhancement for events from western sources and Fe depletion for those from central and eastern sources on the Sun (see Figure 2 in Reames 2014).

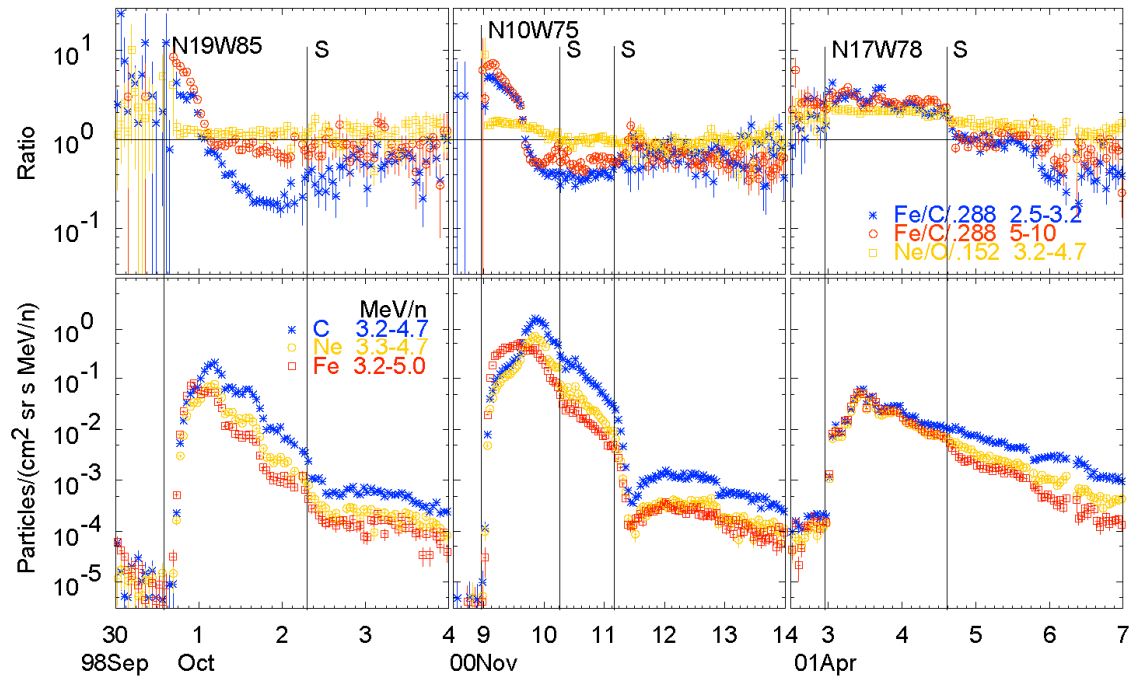

**Fig. 9**. The lower panels show intensities of C, Ne, and Fe *vs.* time for three large gradual events. Upper panels show abundance ratios, normalized to coronal abundances, for Fe/C at two energies, and for Ne/O. Fe/C shows transport-induced enhancement early and depletion later while Ne/O shows much less variation.

These early enhancements in Fe/C or Fe/O were once erroneously considered to be material from an associated flare (*e.g.* by Reames 1990). However, they appear always to lack significant $^3$He enhancements, they occur at all longitudes as seen by multiple spacecraft (Tylka *et al.* 2012) and they are modeled theoretically by transport calculations (Ng Reames, and Tylka 2003). An example of this latter modeling is shown in Figure 10 (see also Mason *et al.* 2006).

**Fig. 10.** Typical theoretical variation of Fe/O by particle transport through self-generated Alfvén waves (from Ng, Reames, and Tylka 2003, 2012). Enhancement or suppression of Fe/O depends upon the shape of the wave spectrum resonant with Fe and with O. That wave spectrum is not really a simple power law, but varies with space and time so Fe/O is not enhanced at very low energies.

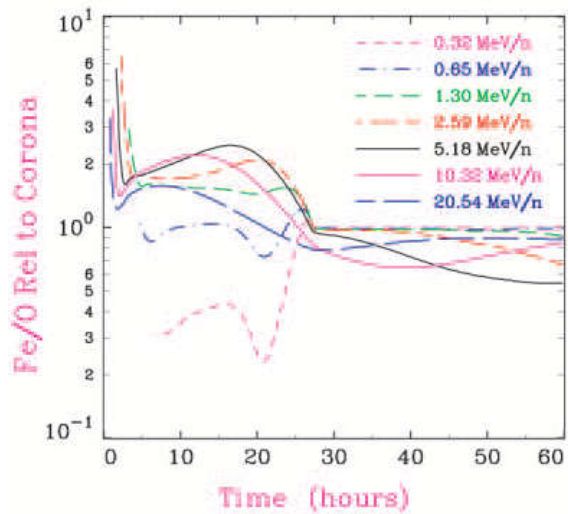

Finally, if we average over a large number of enhanced and depleted regions of Fe/O, we obtain the average gradual-event SEP abundances (Reames 1995a, 1998, 2014). We divide these SEP abundances by the corresponding photospheric abundance from two sources and plot them as a function of the first ionization potential (FIP) of each element as shown in Figure 11. The different photospheric abundance schemes involve the use of different combinations of spectral lines.

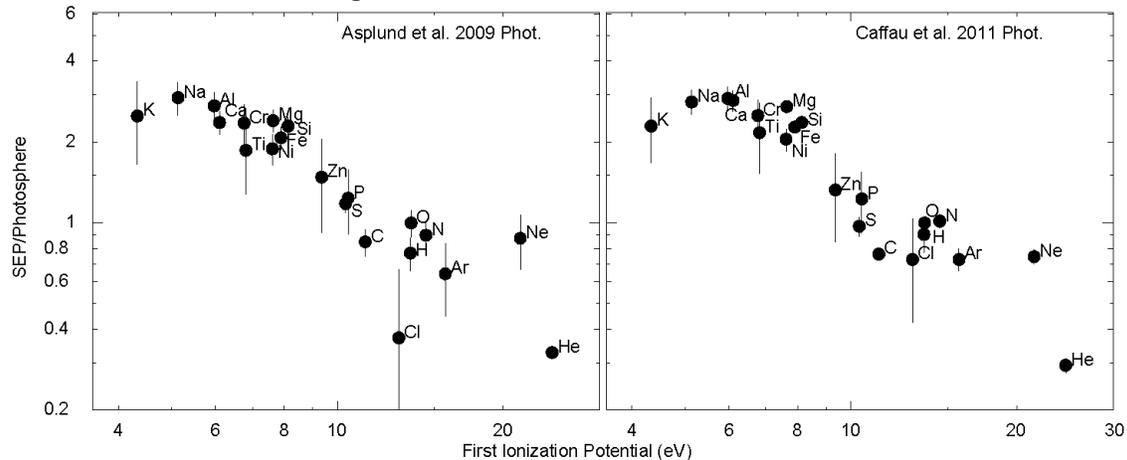

**Fig. 11.** Average element abundances in large gradual SEP events are divided by the corresponding abundances in the solar photosphere as determined by Asplund *et al.* (2009, left panel) and by Caffau *et al* (2011, right panel).

The "FIP effect" shown in Figure 11, known for many years (Meyer 1985), is also seen in coronal spectra and in the solar wind (Schmelz *et al.* 2012) and is caused by differences in the abundances of the corona and photosphere. Low FIP elements are ionized in the photosphere while high-FIP elements are neutral. This affects their transport up into the corona (*e.g.* Laming 2004, 2009). On average, then, shock waves in gradual SEP events merely sample the ambient abundances in the solar corona.

## 3.2 Impulsive Events – Source Temperatures

It has been well known that the events we call impulsive are both $^3$He-rich and Fe-rich (Mason *et al.* 1986; Reames, Meyer, and von Rosenvinge 1994) although the enhancements of $^3$He/$^4$He and Fe/O are not correlated. The EPACT telescope on the Wind spacecraft was designed to extend abundance measurements to the rest of the periodic table above Fe, predicted to be of interest as shown in the left panel of Figure 12 (Reames *et al.* 1990). The right-hand panel in Figure 12 shows the strongly enhanced Z-dependence that was subsequently found (Reames 2000; Reames and Ng 2004) to exceed predictions, much to our surprise. These high-Z enhancements in the 3–10 MeV amu$^{-1}$ region were subsequently confirmed at energies below 1 MeV amu$^{-1}$ by an instrument of completely different design, ACE/ULEIS (Mason *et al.* 2004).

The enhancements in the right-hand panel of Figure 12 do not increase monotonically with Z. In particular, the enhancement in Ne exceeds that in Mg and Si, out of Z order. Since the actual ordering is expected to be a function of *A/Q*, can we find a situation where the *A/Q* values support the observed enhancements? Earlier measurement of abundances in impulsive SEP events led Reames. Meyer, and von Rosenvinge (1994) to suggest that the ion source had a temperature region of 3–5 MK.

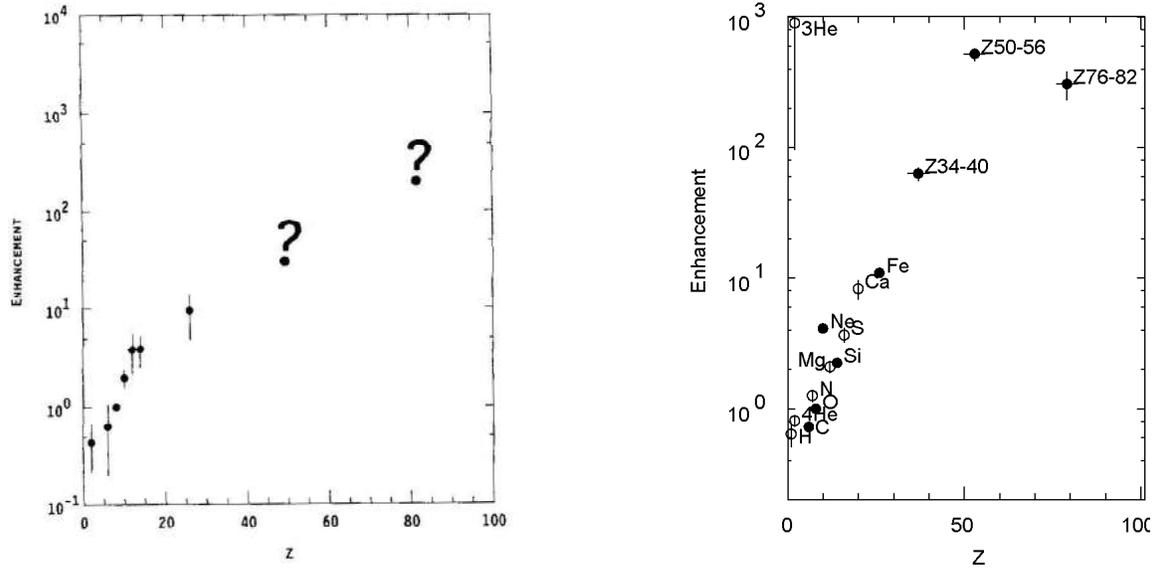

**Fig. 12**. The left panel shows the enhancement of elements throughout the periodic table sought with the design of the Wind/EPACT instrument (Reames *et al.* 1990) while the right panel shows the observations (Reames and Ng 2004).

The theoretical temperature dependence of *A/Q* is shown in Figure 13. These values also agree with the more recent ones up to Fe from Mazzotta et al. (1998). Note that in the red-shaded region *A/Q* for Ne exceeds that for Mg which exceeds that for Si.

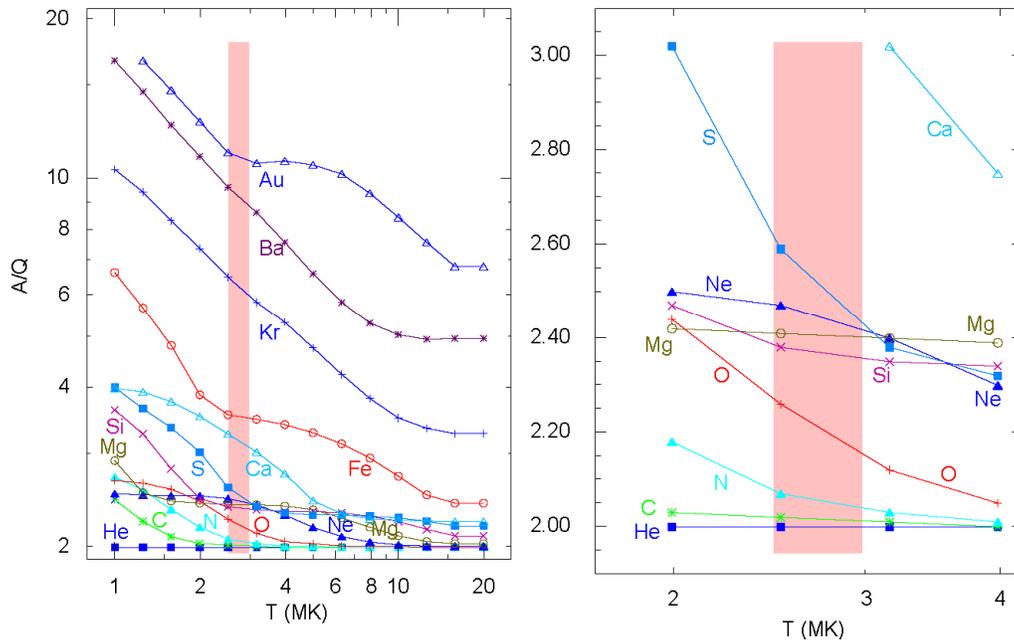

**Fig. 13.** A/Q is shown as a function of equilibrium temperature for several elements (left panel) and enlarged for low Z (right panel). Elements below Fe are from Arnaud and Rothenflug (1985), Fe from Arnaud and Raymond (1992) and sample elements in the high-Z region from Post et al. (1977). The region used for the likely temperature for Fe-rich impulsive SEP events is shaded red (Reames, Cliver, and Kahler 2014a).

Using the average enhancements for 111 impulsive SEP events, defined by their Fe/O enhancements, listed by Reames, Cliver, and Kahler (2014a), and the values of *A/Q* in the shaded region of Figure 13, we plot enhancement *vs.* *A/Q* in Figure 14.

**Fig. 14.** The mean element enhancement in impulsive SEP events (relative to coronal abundances) is shown *vs.* *A/Q* at 2.5–3 MK. The power of the fitted line shown is 3.64±0.15. (Reames, Cliver, and Kahler 2014a)

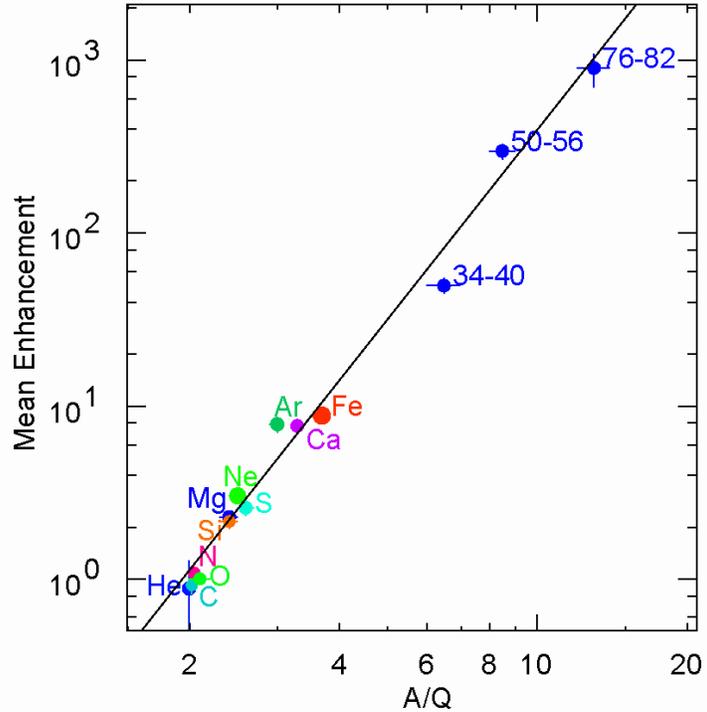

For completeness we compare abundances, relative to O=1000, in impulsive and gradual SEP events for individual elements with expected coronal abundances in Figure 15.

**Fig. 15**. A comparison is shown of the relative photospheric abundances of elements of Asplund *et al.* (2009) corrected to the corona for FIP (light blue), and abundances in gradual (blue) and impulsive (red) SEP events. Individual elements are *not* resolved above Fe in the SEP events and the typical Z resolution is shown along the top of the plot. Even with 20 years of observations it is difficult to populate the highest-Z region in gradual SEP events.

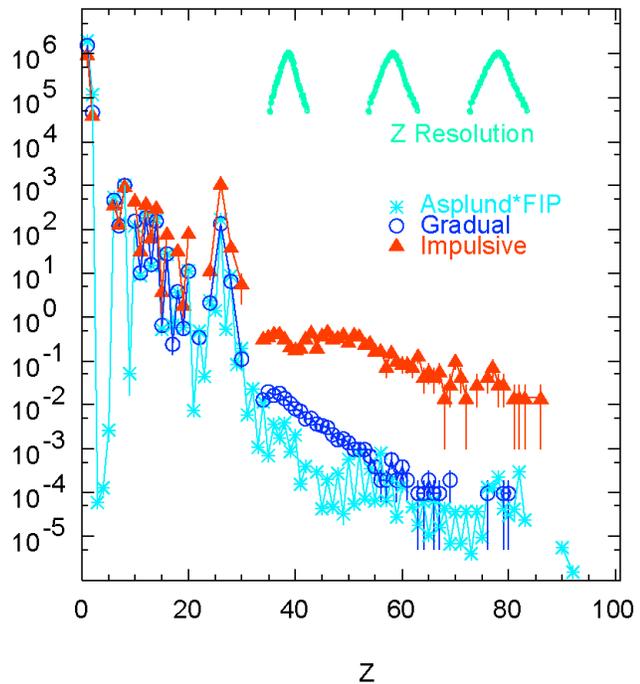

The strong power-law dependence of the abundances in impulsive SEP events has been studied theoretically recently by particle-in-cell simulations of magnetic reconnection regions (Drake *et al.* 2009, 2010; Knizhnik *et al.* 2011; Drake and Swizdak 2014). In this simulation, particles undergo Fermi acceleration as they scatter back and forth against the ends of collapsing islands of magnetic reconnection. Acceleration in impulsive events has previously been associated with the model of solar jets (Shimijo and Shibata 2000) where magnetic reconnection takes place on open field lines (Kahler, Reames, and Sheeley, 2001; Reames, 2002; Archontis and Hood, 2013) in standard or blowout jets (Moore *et al.*, 2010; Kahler, Reames, and Cliver 2015). The open field lines allow the energetic ions to escape directly without passing through material where ionization energy losses could destroy the simple power-law heavy-element enhancements. However the ions may pass through the extremely small amounts of material required for further ionization. Measured ionization states that increase with energy suggest that they have been altered after leaving the source (*e.g.* Klecker *et al.* 2001; DiFabio *et al* 2008).

For comparison we would suggest that the reconnection producing the flaring region beneath a CME involves only closed field lines, preventing particle escape (Reames 2002). Nevertheless, observations of the Doppler-broadened $\gamma$-ray lines from the accelerated "beam" in these large flares suggest that the ions are [3]He-rich (Mandzavidze, Ramaty, and Kozlovsky 1999) and Fe-rich (Murphy *et al.* 1991). Thus the physics of the acceleration in flares and jets may be the same even though the flare-associated ions do not escape these gradual events so we only see the shock-accelerated particles out in space.

For an individual impulsive event, typical variation of the pattern of enhancement *vs. A/Q*, for two values of temperature *T*, is shown in Figure 16. We can fit the pattern for many values of *T*. Plotting the value of $\chi^2$ *vs. T*, we can select the fit and the value of *T* that minimizes $\chi^2$ as done for a large sample of events by Reames, Cliver, and Kahler (2014b). Figure 17 shows $\chi^2$ *vs. T*, for the original fits of 111 impulsive SEP events.

**Fig. 16.** Plot of the enhancements in a single impulsive SEP event *vs. A/Q(T)* and associated fits are for temperatures of 2 MK (red) and 4 MK (blue). Not only does the slope change with temperature, but also the separation of Ca and Fe, of Si and S, or of C and O, for example, changing $\chi^2$ of the fit.

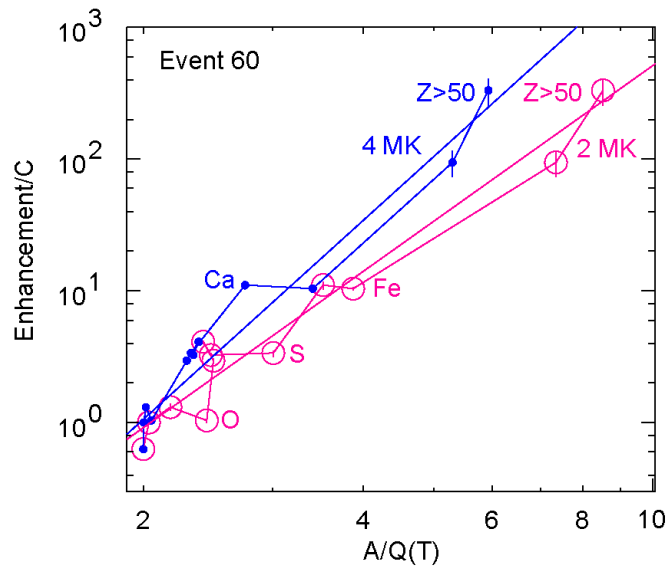

**Fig. 17.** $\chi^2$ is shown as a function of temperature for 111 impulsive SEP events for least-squares fits of abundance enhancements *vs.* *A/Q(T)* in each event (Reames, Cliver, and Kahler, 2014b).

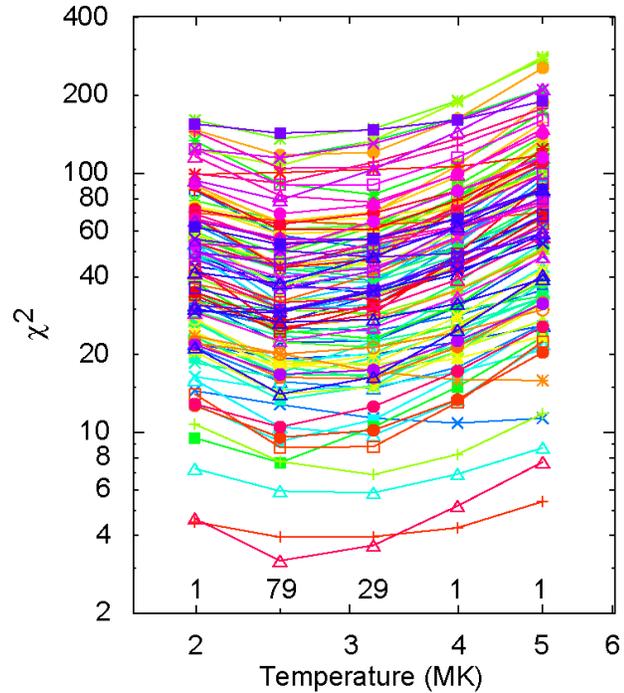

A search for impulsive events with properties that would yield higher and lower *T* (Reames, Cliver, and Kahler 2015) was fairly unproductive. The distribution of the temperature and of the best-fit power of *A/Q* for a refit of all of the impulsive SEP events individually is shown in Figure 18. *Note, however that the SEP source temperatures never approach peak flare temperatures which often rise above 10 MK.*

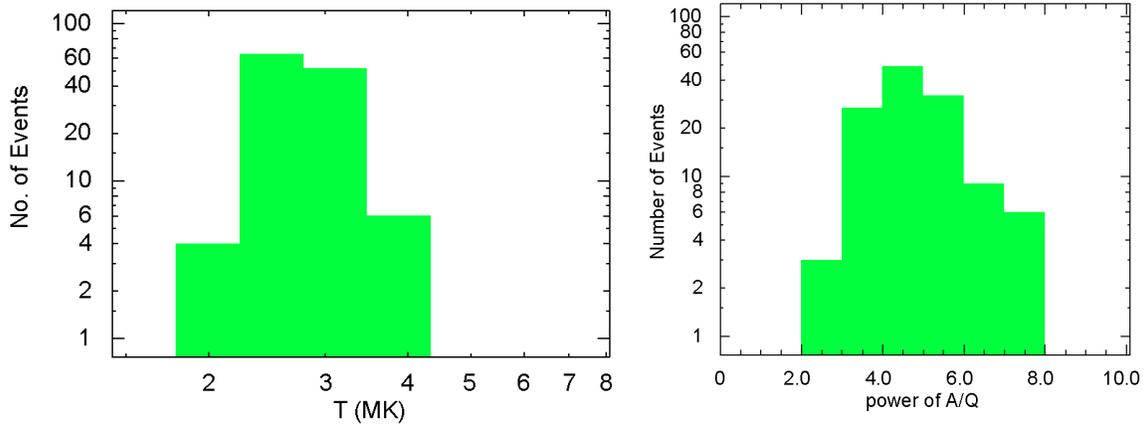

**Fig 18**. The distribution of the temperature (left panel) and of the power of A/Q (right panel) is shown for individual impulsive SEP events (Reames, Cliver, and Kahler 2014b, 2015).

These properties are studied as a function of parameters of the associated X-ray flares and CMEs by Reames, Cliver, and Kahler (2014b, 2015). Generally the correlations are moderate to weak. Figure 19 shows the enhancements, relative to He, as a function of *Z* for individual events with symbols representing the CME width. Nearly 70% of the impulsive SEP events have associated CMEs. Associations with CME speed are somewhat less obvious.

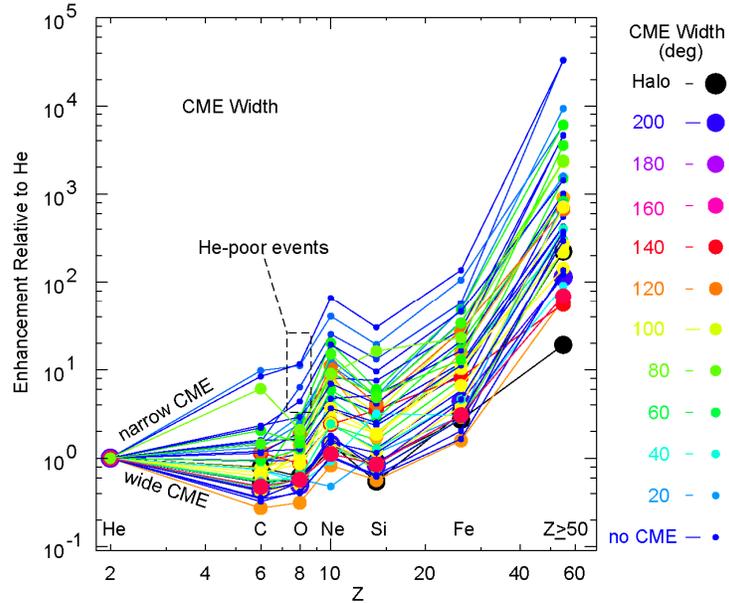

**Fig. 19.** Enhancements, relative to He, are shown *vs. Z* for individual events with heavy elements. Enhancements of individual elements are joined for each event and symbol size and color show the width of the associated CME. Events with greater enhancements associate with narrow CME (Reames, Cliver, and Kahler 2014b)

A comparison of the SEP temperature and the power of *A/Q* for varying soft X-ray class of the associated flare are shown in Figure 20. The left-hand panel shows individual events while the right-hand panel shows averages over X-ray flare class.

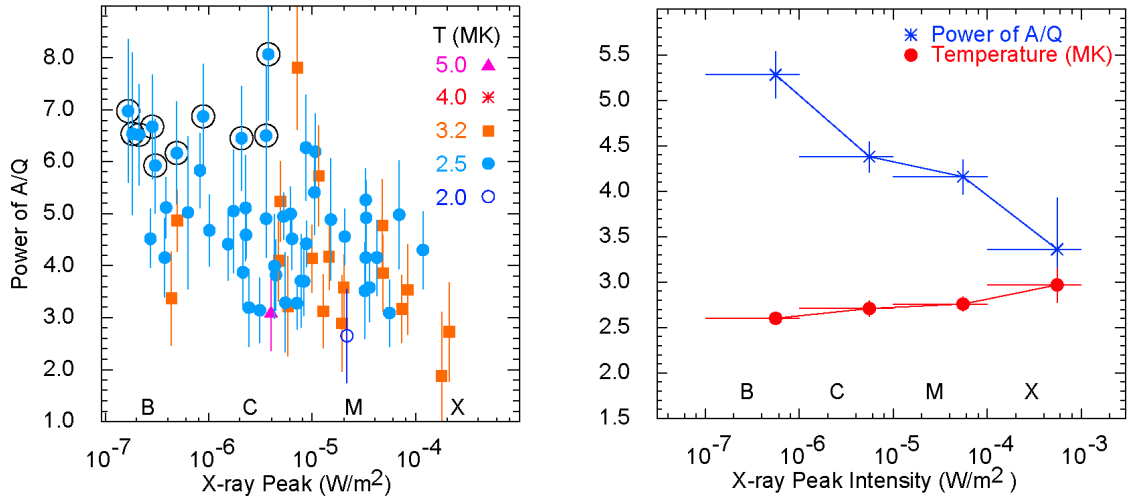

**Fig. 20.** The left panel shows the power of *A/Q vs.* the soft X-ray peak intensity for individual SEP events with temperature as a symbol and color. Circled events are "He-poor" with low He/O The right panel shows variation of the mean temperature and power for each soft X-ray class (after Reames, Cliver, and Kahler 2014b).

If our model involves reconnection on closed loops that contain the energy so as to produce a hot flare and reconnection on nearby open field lines where SEP acceleration occurs, we would expect poor correlation between properties of the flare and the SEPs. What we find is a kind of an inverse big-flare syndrome.

We have focused upon element abundances in impulsive events rather than [3]He. However, we still find that enhancements in [3]He are uncorrelated with those in heavy elements. Nevertheless, the most [3]He-rich events are associated with slow (<700 km s[-1]) narrow (<90°) CMEs and with B- and C-class X-ray flares (Reames, Cliver, and Kahler 2014b).

### 3.3 Gradual Events – Source Temperatures

Recently it has been shown that source plasma temperatures can be extracted from a large number of *gradual* SEP events (Reames 2015). In gradual SEP events, scattering during transport from the shock acceleration near the Sun can produce power-law dependence of element abundances on $A/Q$, enhancements early in the events as weakly scattered ions, like Fe, penetrate the scattering waves, and depressions later where strongly scattered ions, like He, C, or O collect (see Figure 9). *Any process that can produce a power-law dependence upon A/Q can be used to determine source temperatures. Transport can work almost as well as acceleration.* However, a relatively steep power of $A/Q$, positive or negative, is required. A sample analysis of a gradual SEP event is shown in Figure 21

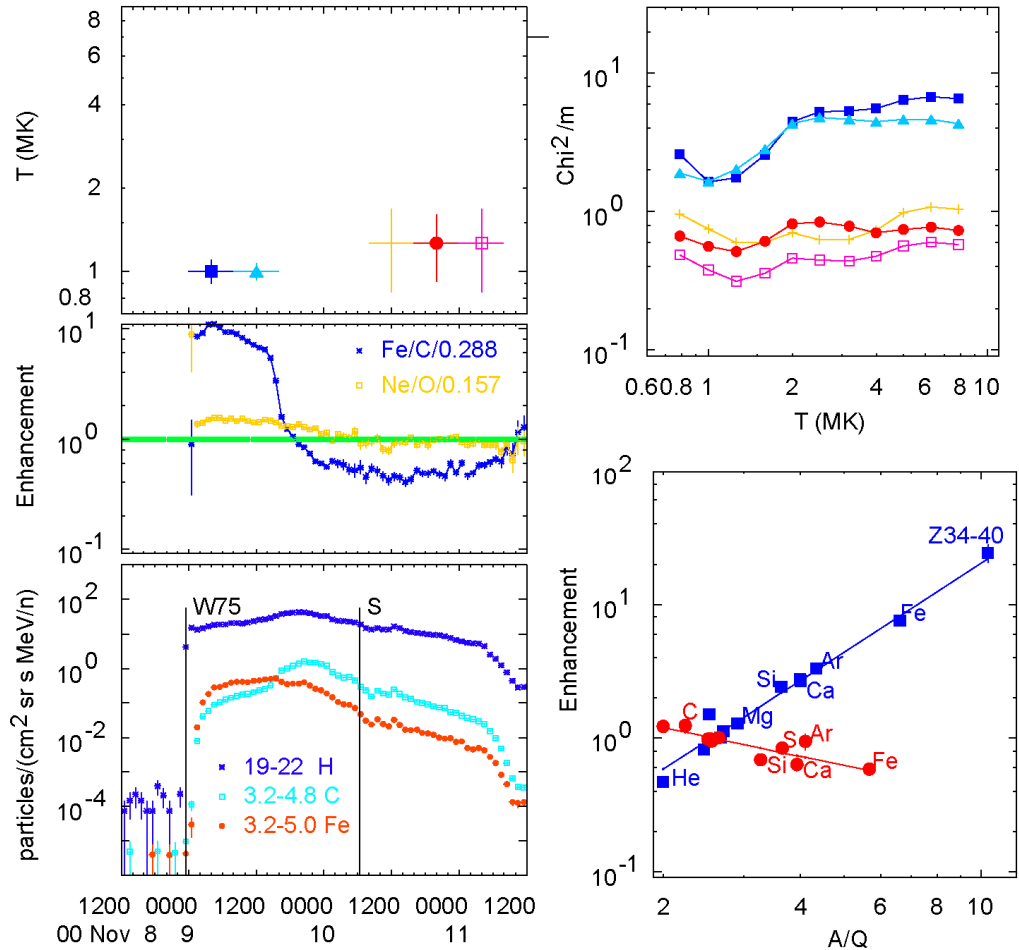

**Fig. 21.** Clockwise from the lower left panel are the intensities of H, C, and Fe during the 8 November 2000 SEP event, the enhancements in Fe and Ne during the event, the best-fit temperatures in color-coded 8-hr intervals, values of $\chi^2/m$ *vs. T*, and best-fit enhancements *vs. A/Q* at two times (Reames 2015).

The center left panel in Figure 21 shows that Fe/C is strongly enhanced early but depressed later in the event. The upper left panel shows temperatures derived during 8-hr periods while the upper right panel shows plots of $\chi^2/m$ *vs. T* corresponding to each time (keyed to the same symbol and color). The variable $m$ is the number of degrees of freedom in the least-squares fit of enhancement *vs. A/Q, i.e.* the number of enhancements measured minus two. The lower right panel shows two of the fits with several elements labeled, the blue squares occur early in the event and the red circles late.

Note that the fits to *the enhancements found early and late in the event have much different slopes, positive and negative, but they give approximately the same source temperature*. Later, the depression of Fe/C is small so the slope is less pronounced so the minimum in $\chi^2/m$ is also more modest and the error in $T$ is larger.

In a study of 45 gradual SEP events with sufficiently strong powers of *A/Q*, Reames (2015) found 69% with temperatures of 0.8–1.6 MK, 24% had impulsive-event source temperatures of 2.5–3.2 MK, possibly from acceleration of impulsive suprathermals by quasi-perpendicular shocks, and there was some evidence of additional highly-ionized material in 24% in the smaller SEP events. Ten GLEs in the sample showed about the same source temperature distribution as the other gradual events.

The events with low source plasma temperatures presumably involve shock waves that sweep up ambient coronal material or suprathermal ions from previous gradual events that do. Figure 22 shows the source plasma temperature *vs.* the associated CME speed. The un-weighted correlation coefficient of -0.49 shows that the seed population for the fastest CMEs tends to originate from ambient coronal plasma, rather than from impulsive suprathermal ions.

**Fig 22**. The average temperature of the source plasma in gradual SEP events is shown *vs.* the speed of the associated CME (Reames 2015).

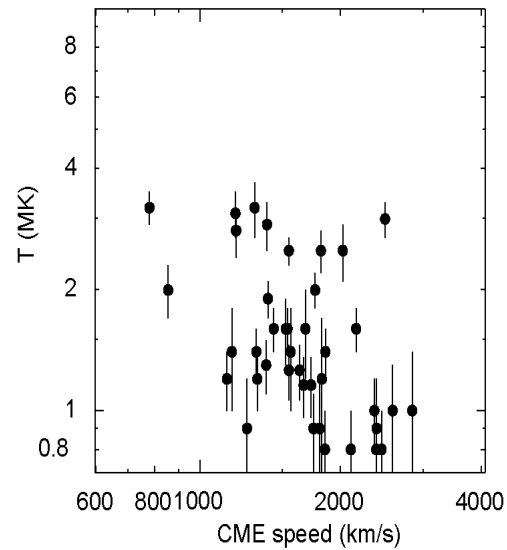

Tylka *et al.* (2005; Tylka and Lee 2006) suggested that shock waves that accelerate impulsive seed material may involve quasi-perpendicular shock waves. For acceleration, a particle downstream must reacquire the shock, a much more demanding task when the magnetic field lies near the shock plane, suggesting a higher threshold energy for acceleration that would preferentially select pre-accelerated ions in the seed population from earlier impulsive (or gradual) SEP events. However, Giacalone (2005) noted that turbulence near the shock would allow oblique shocks better access to lower-energy seed population, especially when $\delta B/B \approx 1$ near the shock and the direction of $B$ becomes irrelevant. Either: i) threshold energies at quasi-perpendicular shock waves with $\delta B/B < 0.5$ favor pre-accelerated ions, or, if not, ii) some CMEs propagate through regions where impulsive suprathermals dominate ambient coronal plasma even at energies of ~10–20 keV/amu .

Gradual SEP events can access a seed population from earlier gradual events. However, there must be an "original" source which is found to involve < 1.6 MK plasma in 64% of the events. So long as the charge states of the ions do not change, reacceleration will not alter the derived temperature.

Physical processes that produce power-law enhancements or suppressions depending upon $A/Q$ are common, and they reward us with a powerful technique to determine the temperature or the ionization pattern of the ions at the original source of these processes. The temperatures help us determine the source, clearly distinguishing impulsive seeds from transport induced enhancements. The events we call impulsive seem to come from heated active-region plasma. Most of the events we call gradual have accelerated ambient coronal material, while some have reaccelerated material of impulsive- or gradual-event origin, or of a more complicated history.

## 4 Discussion

The comparison of "flare" and "shock" acceleration is unfortunate. Shock acceleration evokes a well-defined physical mechanism that has been studied theoretically including self-consistent particle scattering by self-generated waves (*e.g.* Lee 1983, 2005; Jones and Ellison 1991; Ng and Reames 2008; Sandroos and Vainio 2009; Lee, Mewaldt, and Giacalone 2012), but "flare" is not an acceleration *mechanism* at all. Miller *et al.* (1997) summarized possible acceleration mechanisms in flares, to which we may add the newer reconnection modeling of Drake *et al.* (2009). Shock models produce high-energy protons (*e.g.* Zank, Rice, and Wu 2000; Ng and Reames 2008). The flare models that produce enhancements of $^3$He/$^4$He or ($Z$>50)/O *do not* simultaneously produce GeV protons. To understand the sources, we must compare physical mechanisms.

In recent years we have studied element abundances to measure the physics of the formation of the solar corona with the FIP effect or the temperature of the source plasma, parameters of the acceleration, and the properties of related flares and CMEs. But these studies reflect the acceleration physics in the respective shocks and flares, making the origins clearer, especially when we can determine the source plasma temperature. Even studies of the re-acceleration of impulsive flare suprathermal ions by shocks (Mason, Mazur, and Dwyer 1999; Desai *et al.* 2003, 2004; Tylka and Lee 2006; Reames 2015), producing mixed abundances, strengthen the evidence for shock acceleration in these large SEP events. If one were to invoke a flare component for large SEP events like GLEs, then that acceleration mechanism would have to differ from the one for production of $^3$He or Fe-rich events. It would also be one that no one has been able to characterize, one that hides under the cover of the dominant shock acceleration without altering its spectra. No such mystery mechanism is required.

All of the particles we observe must be accelerated on open magnetic field lines. Trapping on flare loops was one problem for the birdcage model since ionization energy losses would modify the low-energy spectra and abundance ratios. That is even more significant for the *A/Q*-dependent impulsive SEPs. In addition, we know that even the impulsive SEPs do not come from the hot flare plasma trapped on magnetic loops.

**Acknowledgement:** This paper was originally presented as an invited plenary talk at the Solar Heliospheric and Interplanetary Environment (SHINE, http://shinecon.org/) conference in Stowe, VT on July 6, 2015. The author thanks Georgia de Nolfo and the conference organizing committee for their support. The author also thanks Steve Kahler for comments on the manuscript and Lun Tan for helpful discussions.